# The nanoscopic submyelin space mediates efficient $K^+$ uptake and shapes nodal action potentials

T. M. Kamsma[1,2], R. van Roij[1], C. Spitoni[2] and M. H. P. Kole[3,4*]

[1]*Institute for Theoretical Physics, Utrecht University, Princetonplein 5, 3584 CC Utrecht, The Netherlands*
[2]*Mathematical Institute, Utrecht University, Budapestlaan 6, 3584 CD Utrecht, The Netherlands*
[3]*Department of Axonal Signalling, Netherlands Institute for Neuroscience (NIN), an Institute of the Royal Netherlands Academy of Arts and Sciences (KNAW), Amsterdam, The Netherlands*
[4]*Cell biology, Neurobiology and Biophysics, Department of Biology, Faculty of Science, Utrecht University, Utrecht, The Netherlands*
*Corresponding author: m.kole@nin.knaw.nl

## Abstract

**Myelin membranes are tightly connected to the axon, forming an electrical route for rapid saltatory conduction of action potentials from node to node. Beneath the myelin, adjacent to the node, lie voltage-gated potassium ($K^+$) channels, whose fundamental role in conduction remains poorly understood. Constrained by biological data, we developed a finite element model of a node of Ranvier encompassing an adaxonal membrane separating oligodendrocyte cytoplasm from the periaxonal space, and solved the full coupled set of physical equations governing electrodiffusion and voltage profiles at the nanoscale. Remarkably, we find that nodal action potentials invert at the inner adaxonal myelin membrane. The hyperpolarization-driven opening of myelinic Kir4.1 channels produces an instantaneous, powerful $K^+$ prebuffering, clamping the periaxonal $K^+$ concentration and enabling high-frequency action potential generation. Together, these findings indicate that the periaxon maintains an electrical feedback loop with the node, and suggest a key evolutionary advantage of juxtaparanodal $K^+$ channel positioning.**

## Introduction

Myelinated axons are one of the most complex cellular developments of the nervous system. The membranes of the oligodendrocytes spirally wrap a multilamellar myelin sheath along the neuronal axon interrupted by micrometer domains of the nodes of Ranvier [1–3]. The tight sealing between the axon membrane and myelin sheath is mediated by transmembrane and adhesion proteins creating nanoscale spaces with distances between 7 to 12 nm, in the paranode and periaxonal domains, respectively [4–6]. Combined, the anatomical pattern creates an electrical low capacitance path along the myelin sheath accelerating the propagation of action potentials (APs) from node to node by a process called saltatory conduction [7,8]. In recent years, evidence accumulated that the nanometer intercellular spaces filled with electrolyte solution also play important roles in the axon-glia metabolic coupling and serve as an activity-dependent signaling pathway [9–11].

Central to both electrical and metabolic functions is the controlled transport and electrodiffusion of ions within the cytoplasm and extracellular solutions. Theoretical and experimental studies have suggested that a single AP causes a ~30 to 100 ms elevation in potassium ($K^+$) concentration [$K^+$], reaching ~1 mM within the periaxonal space of giant invertebrate axons or up to ~100 mM in the periaxonal space of mammalian myelinated axons [12–16]. In myelinated axons this efflux of $K^+$ arises from axonal voltage gated K*v*1 channels which are uniquely clustered to the juxtaparanode domains, beneath the myelin sheath and facing the periaxonal space [12,17]. To maintain electroneutrality the excess of $K^+$ is cleared via a syncytium of oligodendrocytes and astrocytes, starting with the active uptake of $K^+$ by inward-rectifying (Kir) channels, a redistribution via oligodendrocyte and astrocyte gap junctions [18] and $K^+$ ultimately leaving via the blood vessels [19–22]. Interestingly, recent immunogold electron microscopy (EM) showed that the $K^+$ inward-rectifying channel subtype Kir4.1 is localized to the adaxonal membrane, a myelin membrane at the inner site of the myelin sheath facing the periaxonal space [23]. Such channel localization provides a transport route for $K^+$ from the periaxonal space towards oligodendrocyte and astrocyte cytoplasm and suggests a critical role in AP-dependent $K^+$ buffering at the node of Ranvier [9,23]. However, experimentally probing electrodiffusion or performing electrophysiological recordings from the inner myelin membranes is challenging and the $K^+$ flux in these nanometer spaces remains to be determined.

Conventional cable models of myelinated axons have been successful in describing the longitudinal and radial voltage gradients and accurately depict conduction velocity [7,24–28]. At the nanoscale, however, small spaces, such as the periaxon, are associated with greatly fluctuating ion concentrations and the equilibrium potentials are not constant [29]. While cable models can to some extent be adapted to describe ionic dynamics [30,31], they lack the spatial granularity to capture the intricate ionic gradients and coupled transport processes that develop within complex geometries. Finite element (FE) modelling offers a powerful platform for studying electrodiffusion and efficiently solve the ion flow within complex nanoscale geometries, including the nodes of Ranvier [32–35].

Here, leveraging the well characterized node of Ranvier of rodent layer 5 pyramidal neurons in the cerebral cortex [7,26,28,36,37], we developed a FE model that solves the full coupled set of physical ion-transport and electrodiffusion equations in an azimuthally symmetric 3D volume at nanoscale resolution. Importantly, between the periaxonal space and the compact myelin sheath we included a third axial route representing the cytoplasmic tongue, which is connected to the periaxonal space via an adaxonal membrane. Our experimentally constrained and biologically realistic FE model directly solves the underlying physical electrodiffusion equations and yields spatially and temporally detailed solutions for ion concentrations, fluxes, and voltage profiles. We discovered that nodal APs generate inverted (hyperpolarizing) APs at the adaxon membrane. The membrane hyperpolarization prebuffers periaxonal $K^+$ temporally before juxtaparanodal voltage gated $K_V1$ channels are opened. The combined effect is that periaxonal $[K^+]$ is electrodiffusively maintained within the 1–10 mM range, even during high-frequency bursts of action potentials. Finally, we find that the periaxonal space dimension is optimal for activity-dependent $K^+$ clearing, providing an explanatory physical framework for the evolutionary advantage for the anatomical separation of $Na^+$ and $K^+$ channels at the node.

## Results

*FE model and AP generation*

Using the FE-software package COMSOL [38] we created a cylindrically symmetric geometry of a central nervous system (CNS) node of Ranvier constrained by anatomical data and discretized the volume into a fine mesh. The node of Ranvier spans a 2 µm single plasma membrane with flanking paranodes, juxtaparanodes (JXP), and adjacent internode and compartmentalized in the radial direction for the paranodal and periaxonal spaces (**Fig. 1a,b**, **Supplementary Table 1**). Critically, we included a separate compartment of noncompacted oligodendrocyte membranes with a nominal 50 nm intracellular height representing the inner tongue. This compartment contains cytoplasm of the oligodendrocyte and couples to the periaxonal space by a single adaxonal membrane (**Fig. 1c-e**). We numerically solve the Poisson-Nernst-Planck (PNP) equations in a 2D axisymmetric model, revolved into a 3D volume that describe how (charged) species move under combined concentration gradients and electric fields, together with membrane currents described by established Hodgkin-Huxley-type kinetics [39–42].

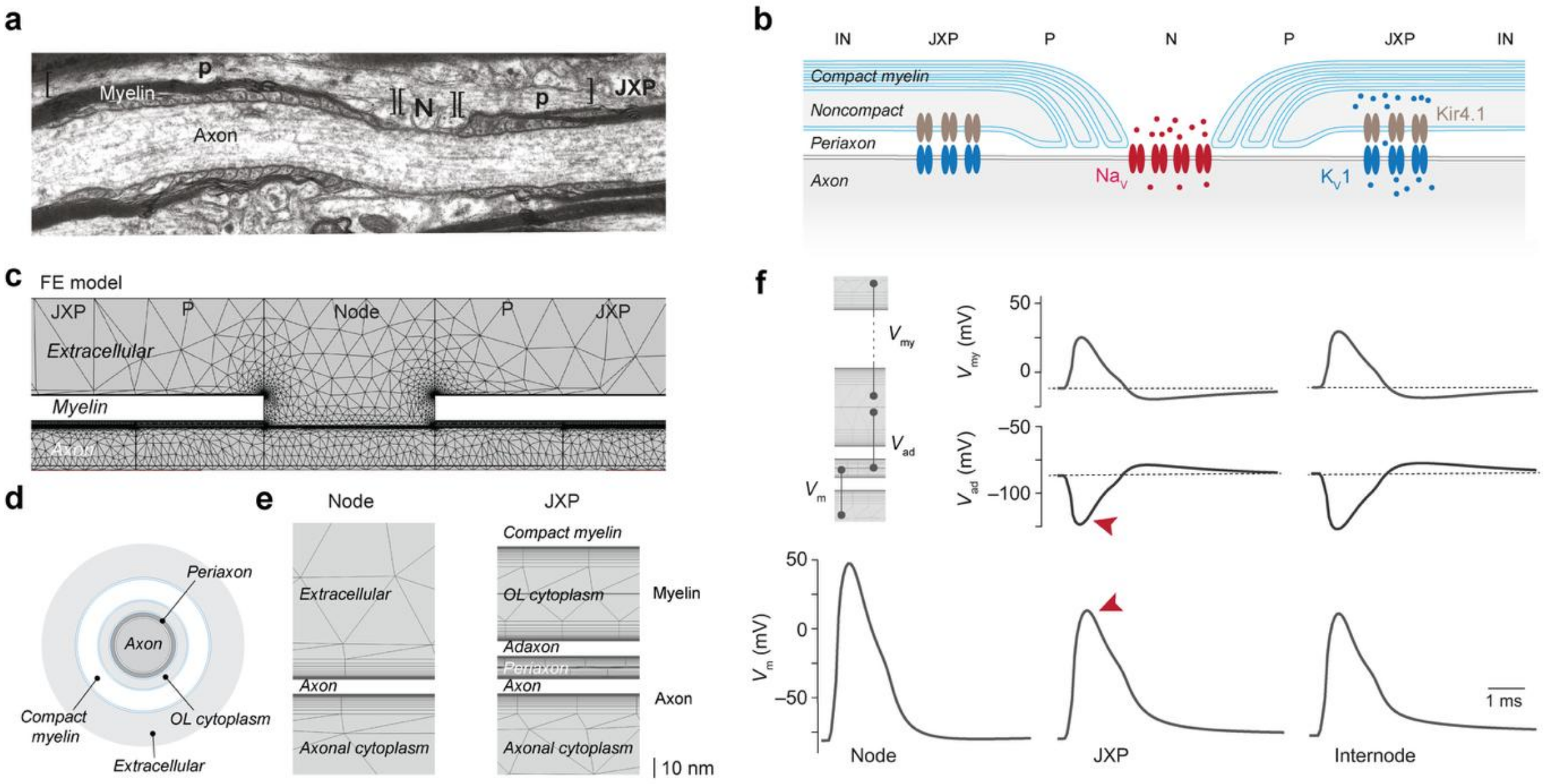

**Fig. 1 │ FE model architecture and simulations of the node of Ranvier**
(**a**) Electron microscope image of a node of Ranvier including the nodal axolemma (N), the paranode (P) and juxtaparanode (JXP). Image adapted from [43].(**b**) Schematic of the myelin and axonal membranes with the ion channels included in the model. Voltage-gated $Na_v$ channels (red) are restricted to the node of Ranvier, the K*v*1 channels (light blue) are localized to the JXP axolemma, and the Kir4.1 channels (dark blue) are within the adaxonal membrane separating the JXP periaxonal space and inner cytoplasmic tongue. (**c**) The COMSOL mesh used for the finite-element model corresponding to the geometry shown in (b). The structure is approximately rotationally symmetric and is revolved around its central axis to obtain a full 3D representation of the node. (**d**) Schematic cross sectional view of the model domains in which PNP was implemented; the extracellular space, oligondendrocyte cytoplasm (inner tongue), the periaxonal space and axon. Cartoon not at scale. (**e**) The zoomed in pictures show the detailed mesh of the COMSOL model and the nanometer changes in ion and voltage profiles that form around the axonal and adonaxol membranes. (**f**) Membrane voltages autonomously generated by FE model simulations at the node, JXP and beginning of the internode. The voltage profiles are the axonal membrane ($V_m$, black), adaxonal $V_{ad}$ ($V_{tongue}$–$V_{periaxon}$, blue) and the transmyelin voltage ($V_{my}$ = $V_{periaxon}$–$V_{extr}$, dark blue). Note the voltage drop in peak amplitude of the propagating $V_m$ and the inversion of the AP at $V_{ad}$ (red arrows).

The reversal potentials are continuously updated according to the dynamic ionic concentrations. We included three ionic species; $K^+$, $Na^+$ and a generic anion, for charge neutrality, with initial ionic concentrations set to typical physiological values (**Supplementary Table 1, Methods**). The concentrations and voltage profiles evolve according to the underlying electrodiffusion PNP equations, while the center of the inner cytoplasmic tongue is fixed at −86 mV [44] (see **Methods**) and the outer boundary of the extracellular space is electrically grounded with fixed ionic concentrations.

To capture the nanoscopic voltage- and concentration profiles that form around the membranes within electrolyte solutions, the computational mesh features a layered sub-nanometer high-resolution structure (**Fig. 1c–e**). Through this mesh the FE model provides a spatially detailed profile of all physical variables throughout all included compartments. In cable theory, the equivalent of this FE model could be a triple cable model (**Supplementary Fig. 1**). However, in the FE model the voltage profiles, and the consequent membrane voltages, as well as the ionic fluxes within each compartment are computed directly from the coupled electrodiffusion and voltage PNP equations. APs were initiated by a brief (0.2 ms) pulse of $Na^+$ influx applied at the outer edge of the axon core (**Supplementary Fig. 2**). Besides this imposed influx, all voltages, ionic fluxes, and concentrations across all compartments evolved according to the underlying physical laws, governed by electrodiffusion and channel equations. As a demonstrative example we show the evolution of the $Na^+$ and $K^+$ concentrations during AP generation in a 3D representation of the model (**Supplementary Movie 1**). The simulations show a ~10 mM increase in axoplasmic $[Na^+]$ to depolarize the membrane to the AP peak amplitude. Interestingly, the increase of axoplasmic $Na^+$ displaces ~3 mM $K^+$ diffusing axially from the nodal into paranodal axoplasm to maintain charge neutrality in the core. Moreover, the juxtaparanode axoplasmic $[K^+]$ decreases further due to the clustered localization of the K*v*1 channels producing radial efflux mediating the AP repolarization.

By plotting the voltage profiles for the different domains, we find that the FE model generated axonal APs with characteristic amplitudes and durations (**Fig. 1f, Supplementary Fig. 2**). The resulting nodal axolemmal voltage waveform showed a voltage threshold of –42 mV, amplitude of 87.9 mV and half-width of ~860 μs, in agreement with experimental voltage imaging or whole-cell recording from the node [7,45], and typically observed with compartmental cable modelling [7,46]. Furthermore, the FE model allows plotting adaxonal membrane potential ($V_{ad} = V_{tongue}–V_{periaxon}$, middle row) and the voltage over the compact myelin and inner tongue ($V_{my}$, top row, **Fig. 1f**). Remarkably, we observed that $V_{ad}$ hyperpolarizes as a result of the increase of the electric potential within the periaxonal space. The peak amplitude of the inverted AP was –116 mV. Together, these results support that the model forms a biophysically accurate representation of the voltage distribution across the membranes around the node of Ranvier and its governing physics.

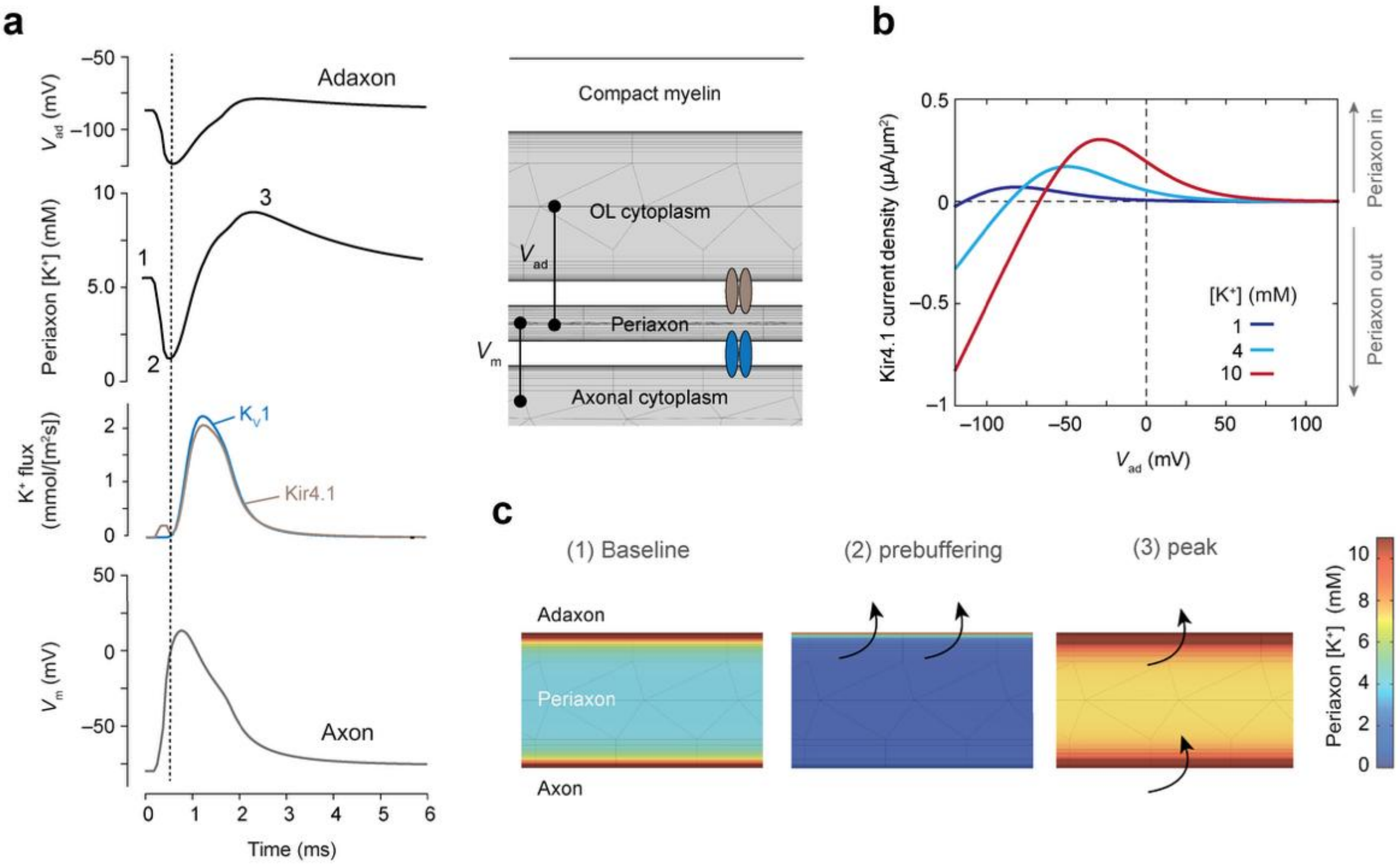

**Fig. 2 | Adaxonal hyperpolarizing APs mediate dynamic and rapid myelin K⁺ uptake**
(**a**) Temporally aligned voltage profiles and K⁺ flux in the JXP. Bottom to top, axonal $V_m$, the total axolemmal K*v*1 K⁺ flux (light blue) overlaid with the adaxonal Kir4.1 K⁺ flux (dark blue), averaged periaxonal [K⁺] (red) with indicated three distinct phases of baseline (1), prebuffering (2) and peak (3) stages, $V_{ad}$ hyperpolarization. Dotted line indicates the peak of the prebuffering. Schematic of the JXP domain indicating the channel localization membrane and voltages plotted in a. (**b**) K⁺ current density–voltage relationship of the Kir4.1 model for distinct extracellular K⁺ concentrations and varying adaxonal membrane potentials. For illustration purposes current direction is plotted relative to the periaxon. Out, K⁺ flowing out from periaxon into OL cytoplasm. In, K⁺ flowing from OL cytoplasm into periaxon. (**c**) Zoom of color-coded periaxonal space [K⁺] during three phases of the AP indicated in (a). Note the FE model reveals the high ionic concentrations near the plasma membrane walls according to the physical Debye length constants. Arrows illustrate Kir4.1-mediated K⁺ uptake into oligodendrocyte cytoplasm (top) and K*v*1-mediated influx from the axoplasm (bottom).

### *Periaxonal K⁺ accumulation in the JXP*

Next, we examined the axon-myelin K⁺ dynamics associated with APs in the JXP. Consistent with the localization of K*v*1 channels and ~80 mV large APs the simulations showed significant K⁺ influx into the periaxonal space during repolarization (**Fig. 2a**). When reviewing the periaxonal [K⁺] transient we observed a remarkable temporal sequence. Within the first millisecond the periaxonal [K⁺] is nearly depleted, from an average 5.5 to 1.3 mM, followed by an increase to 8.9 mM, peaking at 2.3 ms, and a return to the baseline concentration at ~6 ms. To better understand the spatiotemporal dynamics of the K⁺ ions in the space we also plot the average K*v*1 and Kir4.1 fluxes. Instantaneously with the rising phase of the nodal AP $V_{ad}$ hyperpolarizes, due to the passive propagation of the AP and subsequent voltage increase within the periaxonal space (**Fig. 1d, 2a**). The hyperpolarization of $V_{ad}$ can be explained by local circuit currents associated with the initial Na⁺ entry. The capacitively coupled axon and adaxon membranes, in combination with the high periaxonal axial resistance, depolarizes the electrolyte solution in the periaxonal space (**Supplementary Fig. 3**).

Hyperpolarization of $V_{ad}$ greatly increases the driving force for $K^+$ and thereby promotes $K^+$ efflux through the voltage- and concentration-dependent Kir4.1 channels (**Fig. 2a-c**). Since the axonal K*v*1 channels open with a voltage- and time-dependent delay of several hundred microseconds [41], $K^+$ entry into the periaxon is delayed and the efflux mediated by the increased electrical driving force causes a net depletion to a low concentration of 1.3 mM, a phenomenon we term '*$K^+$ prebuffering*'. This transient $K^+$ prebuffering leads to a decrease in $K^+$ concentration, rapidly limiting the Kir4.1-mediated flux (**Fig. 2a-c**). About 200 μs later K*v*1 channels open to produce a periaxonal [$K^+$] influx, which is efficiently buffered by the cytoplasmic tongue due to the starting $V_{ad}$ of –116 mV. The periaxonal [$K^+$] peaks at ~9 mM at 3 milliseconds after the AP onset and relaxes fully back toward the baseline state several milliseconds later (**Fig. 2a**).

*High-frequency action potentials*

The FE model simulations show that the rise in periaxonal $K^+$ is effectively shunted within a few milliseconds for a single AP. To explore the parameter range of periaxonal [$K^+$] we simulated physiologically relevant high firing frequencies (>100 Hz) when extracellular $K^+$ steeply rises around the axosomatic membrane of layer 5 pyramidal neurons [26,44]. In the FE model we generated a burst of 4 APs at 150 Hz, by repetitively injecting $Na^+$ pulses and compared electrodiffusion with and without adaxonal Kir4.1 membrane conductance (5 and 0 nS $\mu m^{-2}$, respectively). **Figure 3** shows that, with the control Kir4.1 density, the node of Ranvier is capable to sustain APs at 150 Hz, with each AP within the train reaching starting near the resting potential and reaching peak amplitudes of at least ~ +36 mV. The periaxonal [$K^+$] fluctuates between 1.3 and 9.0 mM without increasing the baseline [$K^+$]. In contrast, without Kir4.1 channels $K^+$ accumulates within the nano-confined periaxonal space peaking at an order of magnitude higher during the burst (~80 mM, **Fig. 3b**). These concentrations are in range of the upper limit which were mathematically predicted on the basis of electrical intra-axonal recordings [13]. While the nodal $V_m$ rises normally for the first AP it fails to repolarize and the remaining sharp voltage transients reflect the externally injected $Na^+$ without APs (**Fig. 3b**). Together, these model simulations indicate that the submyelin space actively contributes to, and is required for, the repolarization of the nodal domain.

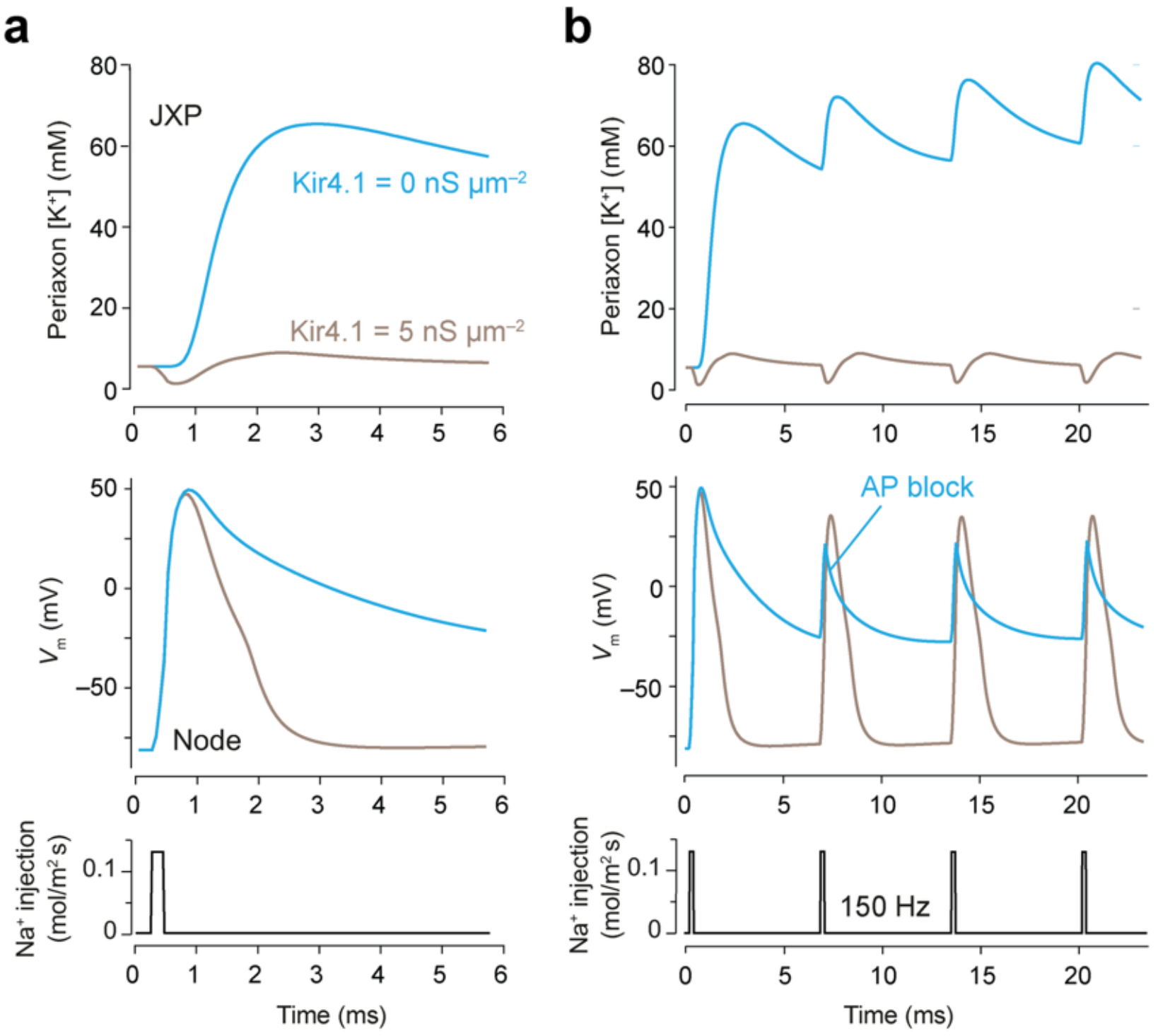

**Fig. 3 │Periaxonal space $K^+$ efflux repolarizes the nodal AP**
(**a**) AP model simulations with (back) and without (blue) Kir4.1 channels in the adaxonal membrane. Bottom shows the $Na^+$ injection protocol to evoke a nodal AP (middle), and the related periaxonal $[K^+]$ (top). (**b**) same FE simulations extended for 25 seconds to record an AP burst evoked by a 150 Hz $Na^+$ injection protocol. The subsequent APs within the burst fail from the second AP when myelin $K^+$ buffering is removed (blue).

### *Periaxonal space optimal for $K^+$ clearance*

The height of the periaxonal space, as schematically indicated in **Fig. 4a,** is set to ~12 nm by transmembrane proteins like myelin-associated glycoprotein [5,6]. Previous experimental studies and cable modelling indicated that the space height limits the conduction velocity and increases when the periaxonal space is narrower [7,47]. How the periaxonal space shapes the ion dynamics of $K^+$ is unknown. In analogy, neuronal synaptic clefts have an optimal dimension for synaptic strength due to a tradeoff between the resistance of the extracellular space and the neurotransmitter concentration profile [48]. Since the Kir4.1 channel is gated by both $V_m$ and $[K^+]$ the intercellular height may have similar complex implications for $K^+$ uptake. With a smaller volume periaxonal $K^+$ concentrations are expected to increase and uptake may be facilitated [40,49]. To explore this, we ran FE model simulations with periaxonal space heights ($\delta_{pa}$) between 7.4 nm, similar to the narrow space dimensions at the paranodes [50], and a maximum of 400 nm. The results show that with larger $\delta_{pa}$ the axolemma $V_m$ increases in peak amplitudes consistent with cable models [7]. In contrast, the peak hyperpolarization of $V_{ad}$ becomes smaller with increasing $\delta_{pa}$ (**Fig. 4b**).

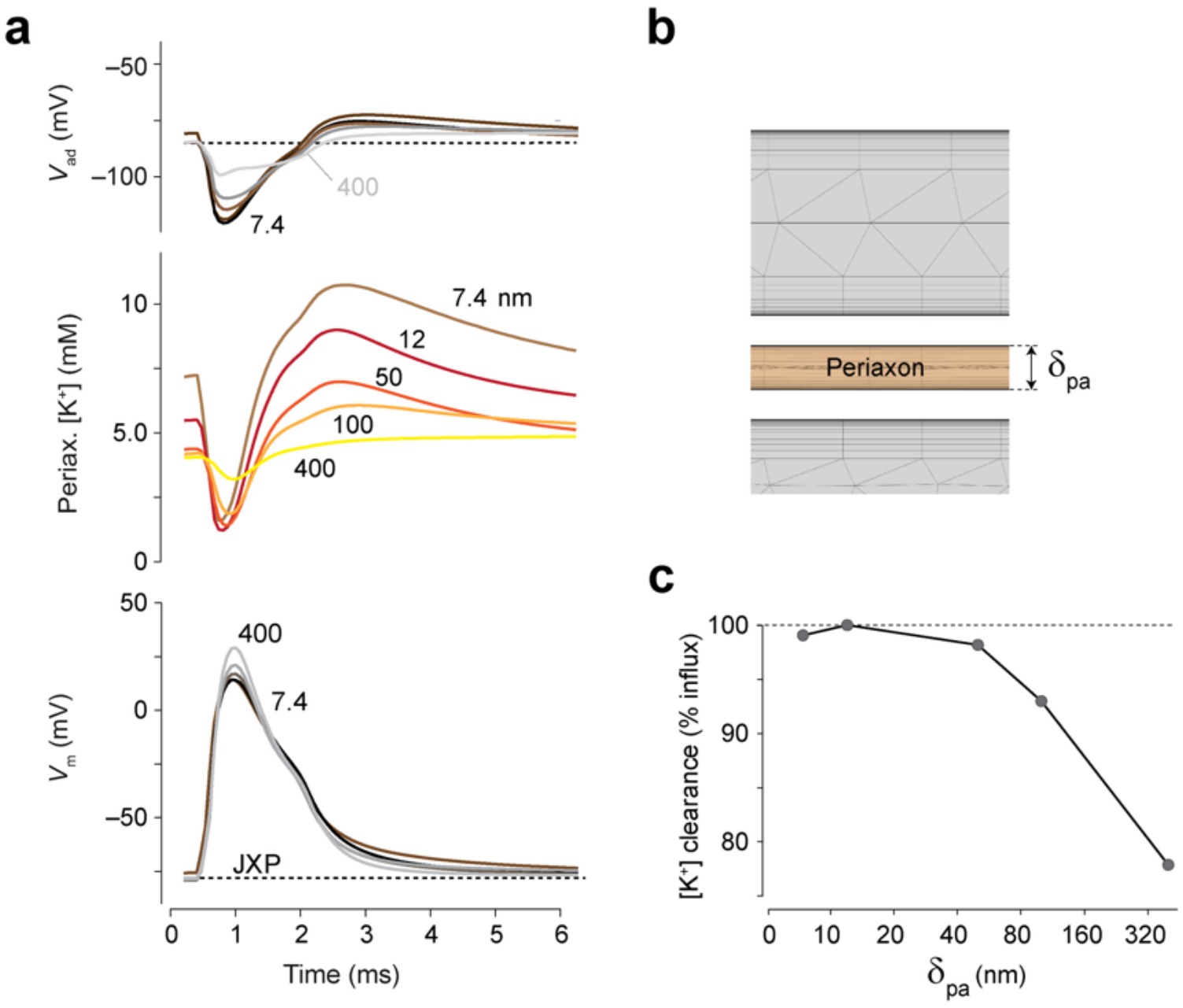

**Fig. 4│ Periaxonal nanoscale dimensions optimal for $K^+$ buffering**

(**a**) Top to bottom; $V_{ad}$, periaxonal [$K^+$] and axonal $V_m$ for different periaxonal heights from 400 to 7.4 nm. Note the larger voltage drop of the AP amplitude for $V_{ad}$ compared to $V_m$ with the larger spacing. (**b**) COMSOL mesh schematic with periaxonal space which is varied in its height. (**c**) percentage of total periaxonal K$v$1-mediated $K^+$ influx ($C_{Kv1}$) that is cleared through adaxonal Kir4.1 channels ($C_{Kir}$) as a function of $\delta_{pa}$.

The ability of Kir4.1 channels to clear $K^+$ depends on this intricate interplay between $\delta_{pa}$, periaxonal [$K^+$] and $V_{ad}$. This is compiled in **Fig. 4c**, where we calculate the percentage of K$v$1-mediated $K^+$ influx cleared through adaxonal Kir4.1 channels. With the biologically realistic dimension of ~12 nm, the influx $C_{Kv1}$ from JXP K$v$1 channels is near-perfectly balanced by Kir4.1-mediated efflux $C_{Kir}$, achieving 100% for a single AP, effectively, fully clearing $K^+$. For larger $\delta_{pa}$ dimensions the clearing efficacy decreases. The explanation behind this can be found in the differences of adaxonal membrane voltage and periaxonal [$K^+$], shown in **Fig. 4**. During hyperpolarization of the $V_{ad}$, the comparatively increased Kir4.1 activity due to a smaller $K^+$ concentration decrease during prebuffering for larger heights, is counteracted by the decreased electric driving force due the lower adaxonal hyperpolarization. During AP repolarization voltage amplitudes differ only marginally, while $K^+$ concentrations differ almost two-fold, meaning a significantly higher Kir4.1 activity for smaller periaxonal heights. However, below ~10 nm this balance shifts, as a slower repolarization of the axonal $V_m$ occurs, due to the increased baseline periaxonal [$K^+$], resulting in a higher sustained $K^+$ influx. Consequently, a slight loss of uptake efficiency is also found for a periaxonal height of 7.4 nm. This is possibly compounded by the larger role in narrower spaces of the inhomogeneous [$K^+$] profiles, which radially increase near the membranes, similar to the electric potentials. This effect is explained by the electrostatic effect on the charges which distribute with the Debye length of approximately ~1 nm [28]. As the $\delta_{pa}$ is decreased, these charge gradients represent a more significant fraction of the volume, meaning that the average periaxonal space $K^+$ concentration becomes elevated (**Fig. 4a, b**). Together, these findings indicate that the periaxonal height between ~10–20 nm is optimal for both $K^+$ clearance and nodal AP repolarization.

## Discussion

By solving the full set of electrodiffusion equations in an anatomically realistic geometry of the node of Ranvier we find that the nanoscale dimensions of the periaxonal volume and inner cytoplasmic tongue create ionic profiles and voltage dynamics that strongly couple to the electrical field of the node. A key discovery is that micrometers up- and downstream from a node the redistribution of capacitive currents rapidly increases the periaxonal space voltage, causing hyperpolarization of the adaxonal membrane, which facilitates the efflux of $K^+$ into the myelin sheath cytoplasmic tongue.

The magnitude of the adaxonal membrane change (~30 mV in amplitude) and its direction (hyperpolarizing) may seem counter-intuitive but can be understood by the cable theory of ephaptic coupling of closely apposed membranes [51,52]. Propagating APs generate a spatial wave characterized by a local current sink in the extracellular field at the depolarizing site but at the adjacent sites, both in the wake and the front, show an opposite sign in the extracellular field since charges rapidly distribute longitudinally via the cytoplasm to leave the axon radially and create a local circuit [52]. At the node of Ranvier, the circuit loops are extremely polarized; a very high density of positive currents radially enters the axoplasm producing a large axial current flow charging adjacent para- and juxtaparanode cytoplasm. Consistent with the capacitive coupling theory, our simulations show that the resistance of periaxonal space shapes the amplitude of $V_{ad}$ hyperpolarization (**Fig. 4**).

To obtain direct experimental evidence for the capacitive coupling to the node electrical recording from the adaxonal myelin membrane are required. While such experiments are not feasible these may become attainable in the future when employing high-speed voltage imaging. Recent developments show that voltage sensors can be successfully expressed in the compact membranes of the myelin sheath [53]. In that study voltage imaging at a low speed of 0.1 kHz revealed slow depolarization of myelin abolished by the genetic deletion of oligodendroglial-specific Kir4.1 channels [53]. These data are consistent with the present model showing myelinic Kir4.1 is a major component of neuronal activity-dependent $K^+$ clearance. Furthermore, our FE electrodiffusion simulations without Kir4.1 conductance afford an explanation for the pathological changes in oligodendroglia-specific Kir4.1 knockout mouse, exhibiting impaired activity-dependent $K^+$ uptake and epileptic seizures [9,23,44,54]. Numerical estimations based on van 't Hoff relation predict that the pressure within the periaxonal space for the large ~80 mM $K^+$ concentrations could increase by more than 200 kPa (26 kPa per sustained 10 mM increase), resulting in swelling of several nm for compact myelin, or even tens of nm for a single membrane. The predicted broadening of the nodal AP may act to evoke AP initiation locally in hyperexcitable nodal domains and trigger ectopic APs, as was shown recently [53]. Future optical recording from single adaxonal membranes, however, will require constructs expressing voltage sensors not only with oligodendrocyte-specific promoters but also targeting adaxonal protein motifs and optimize the light collection to acquire AP transients at >10 kHz.

The present study sheds light on the long-standing fundamental question why K*v*1 channels are ‘juxta’ from the nodal $Na^+$ channels and under the myelin sheath [12,55]. Given the narrow space for ionic currents, it is generally believed that JXP K*v*1 channels are not activated in normally myelinated axons but instead provide a safety factor for re-excitation of the node when paranodal membranes are detached from the axolemma [56]. However, genetic deletion experiments preventing the clustering of K*v*1 channel isoforms to the JXP shows evidence for a role in setting the refractory period of APs in baseline conditions [57]. Such direct electrical contribution to membrane excitability and repolarization of the node is consistent with the FE electrodiffusion model (**Fig. 3**). However, an electrical contribution of K*v*1-mediated repolarization can also be achieved by clustering K*v*1 with Na*v* channels directly within the nodal axolemma. What is the benefit of clustering K*v*1 adjacent to the $Na^+$ channels? Based on the increased efficacy of $K^+$ uptake we speculate that the metabolic role is one of the key drivers for this evolutionary development. We find that the myelin [$K^+$] uptake via Kir4.1 greatly benefits from the capacitive hyperpolarization of the adaxon which is optimal at micrometers distance from the $Na^+$ enriched sites and the spatial separation facilitates the [$K^+$] uptake being instantaneous and temporally coupled to the upstroke of the nodal AP. In contrast, at the $Na^+$ rich plasma membrane the electrical field charges the extracellular solution in opposite directions and will depolarize nearby membranes. Furthermore, solving the 3D electrodiffusion indicates a maximum $K^+$ clearance with the observed space (10–20 nm), being a trade-off between concentration profiles, longitudinal resistance and the adaxonal hyperpolarization.

The predicted optimal periaxonal space size is remarkably similar to the optimal synaptic cleft space for synaptic transmission and activation of ionotropic receptors [48,58], and generally supports the idea that the nanometer axon-myelin spaces act as a biochemical signaling compartments [10,11]. These findings are also in keeping with the contribution of electrical fields in small neuronal or glial spaces, such as neuronal synaptic clefts [29,48]. Notably, the evolutionary origin of glia ensheathment of neuronal membranes, already present in bilaterian invertebrates, preceded myelin compaction and plays a critical role in the trophic support of axons [59,60], which in part is mediated by the oligodendroglial Kir4.1 (Ref. [9]).

The axon-glia membrane architecture and routes for axon/myelin $K^+$ release and uptake are more complex than simulated in the present FE model. It involves a multicellular syncytium and other $K^+$ channels contributing to local $K^+$ rise including the nodal non-inactivating K*v*7 channels, myelinic gap junctions, hemichannels and astrocyte-oligodendrocyte coupling [18,19,37,61] and within the myelin sheath K*v*1-containing cytoplasmic loops are spirally wrapped along the axis to connect opposing juxtaparanodal domains [1,19,22]. Furthermore, the adaxonal membrane also contains glutamatergic receptors [9–11]. However, the present multiphysics model of the node of Ranvier offers a wide range of possibilities to study these proteins. For example, the modeling framework can be extended to explore mechanisms for each of these transmitter or ion channels and make quantitative predictions about their functional roles. Additionally, the simulations can be straightforwardly extended to include chemical reactivity, fluid flow, and pressure, meaning the current multiphysics framework could allow to solve and explore the contribution of the membrane mechanics to the AP voltage waveform. For example, one remarkable and recent discovery is the presence of two-pore domain $K^+$ channels TREK and TRAAK within the node of Ranvier, clustered with the $Na^+$ channels [45,62,63]. The mechanoreceptor channels transduce mechanical changes of the lipid membrane bilayer into ionic currents, opening with pressure changes up to 100 mmHg (~13.3 kPa) [64], in line with our forementioned numerical estimates in the FE modelling for the nodal membrane (see Online Methods for details) and the simulated ion concentrations to produce the APs.

The new modeling framework thus opens the possibility of studying the mechanical fluid dynamics contributing to AP repolarization, offering an exciting and important avenue for the exploration of the fundamental features of biologically complex systems to expand our mechanistic insights into the physics and nanopysiology of axon-myelin architectures.

## References

1. Jarjour, A. A. *et al.* The formation of paranodal spirals at the ends of CNS myelin sheaths requires the planar polarity protein Vangl2. *Glia* **68**, 1840–1858 (2020).
2. Rasband, M. N. & Peles, E. Mechanisms of node of Ranvier assembly. *Nat. Rev. Neurosci.* **22**, 7–20 (2021).
3. Hirano, A. & Dembitzer, H. M. The transverse bands as a means of access to the periaxonal space of the central myelinated nerve fiber. *J. Ultrastruct. Res.* **28**, 141–149 (1969).
4. Djannatian, M. *et al.* Two adhesive systems cooperatively regulate axon ensheathment and myelin growth in the CNS. *Nature communications* **10**, 4794 (2019).
5. Pronker, M. F. *et al.* Structural basis of myelin-associated glycoprotein adhesion and signalling. *Nature communications* **7**, 13584 (2016).
6. Li, C. *et al.* Myelination in the absence of myelin-associated glycoprotein. *Nature* **369**, 747–750 (1994).
7. Cohen, C. C. H. *et al.* Saltatory Conduction along Myelinated Axons Involves a Periaxonal Nanocircuit. *Cell* **180**, 311-322.e15 (2020).
8. Huxley, A. F. & Stämpfli, R. Evidence for saltatory conduction in peripheral myelinated nerve fibres. *J Physiology* **108**, 315–339 (1949).
9. Looser, Z. J. *et al.* Oligodendrocyte–axon metabolic coupling is mediated by extracellular K+ and maintains axonal health. *Nat. Neurosci.* **27**, 433–448 (2024).
10. Micu, I., Plemel, J. R., Caprariello, A. V., Nave, K.-A. & Stys, P. K. Axo-myelinic neurotransmission: a novel mode of cell signalling in the central nervous system. *Nature Reviews Neuroscience* **19**, 49–58 (2018).
11. Micu, I. *et al.* The molecular physiology of the axo-myelinic synapse. *Experimental neurology* **276**, 41–50 (2016).
12. Chiu, S. Y. & Ritchie, J. M. Evidence for the presence of potassium channels in the paranodal region of acutely demyelinated mammalian single nerve fibres. *J. Physiol.* **313**, 415–437 (1981).
13. David, G., Barrett, J. N. & Barrett, E. F. Evidence that action potentials activate an internodal potassium conductance in lizard myelinated axons. *J. Physiol.* **445**, 277–301 (1992).
14. Chiu, S. Y. Functions and distribution of voltage-gated sodium and potassium channels in mammalian schwann cells. *Glia* **4**, 541–558 (1991).
15. Frankenhaeuser, B. & Hodgkin, A. L. The after-effects of impulses in the giant nerve fibres of Loligo. *J. Physiol.* **131**, 341–376 (1956).
16. Adelman, W. J., Palti, Y. & Senft, J. P. Potassium ion accumulation in a periaxonal space and its effect on the measurement of membrane potassium ion conductance. *J. Membr. Biol.* **13**, 387–410 (1973).
17. Rasband, M. N. *et al.* Potassium channel distribution, clustering, and function in remyelinating rat axons. *The Journal of neuroscience* **18**, 36–47 (1998).
18. Menichella, D. M. *et al.* Genetic and Physiological Evidence That Oligodendrocyte Gap Junctions Contribute to Spatial Buffering of Potassium Released during Neuronal Activity. *The Journal of neuroscience* **26**, 10984–10991 (2006).
19. Kamasawa, N. *et al.* Connexin-47 and connexin-32 in gap junctions of oligodendrocyte somata, myelin sheaths, paranodal loops and Schmidt-Lanterman incisures: implications for ionic homeostasis and potassium siphoning. *Neuroscience* **136**, 65–86 (2005).

20. Kofuji, P. *et al.* Kir potassium channel subunit expression in retinal glial cells: Implications for spatial potassium buffering. *Glia* **39**, 292–303 (2002).

21. Kofuji, P. & Newman, E. A. Potassium buffering in the central nervous system. *Neuroscience* **129**, 1043–1054 (2004).

22. Rash, J. E. *et al.* KV1 channels identified in rodent myelinated axons, linked to Cx29 in innermost myelin: support for electrically active myelin in mammalian saltatory conduction. *Journal of Neurophysiology* **115**, 1836–1859 (2016).

23. Schirmer, L. *et al.* Oligodendrocyte-encoded Kir4.1 function is required for axonal integrity. *eLife* **7**, e36428 (2018).

24. Rall, W. *Core Conductor Theory and Cable Properties of Neurons. In "The Nervous System"(ER Kandel, Ed.), Vol. 1, Part 7*. (1977). doi:10.1002/cphy.cp010103.

25. McIntyre, C. C., Richardson, A. G. & Grill, W. M. Modeling the Excitability of Mammalian Nerve Fibers: Influence of Afterpotentials on the Recovery Cycle. *J Neurophysiol* **87**, 995–1006 (2002).

26. Jamann, N. *et al.* Layer 5 myelination gates corticothalamic coincidence detection. *Nat. Commun.* **16**, 10922 (2025).

27. Gow, A. & Devaux, J. A model of tight junction function in central nervous system myelinated axons. *Neuron glia biology* **4**, 307–317 (2008).

28. Arancibia-Carcamo, I. L. *et al.* Node of Ranvier length as a potential regulator of myelinated axon conduction speed. *eLife* **6**, (2017).

29. Savtchenko, L. P., Poo, M. M. & Rusakov, D. A. Electrodiffusion phenomena in neuroscience: a neglected companion. *Nat. Rev. Neurosci.* **18**, 598–612 (2017).

30. McDougal, R. A., Hines, M. L. & Lytton, W. W. Reaction-diffusion in the NEURON simulator. *Front. Neuroinformatics* **7**, 28 (2013).

31. Carnevale, N. T. & Hines, M. L. *The NEURON Book*. (Cambridge University Press, 2006).

32. Dione, I., Deteix, J., Briffard, T., Chamberland, E. & Doyon, N. Improved Simulation of Electrodiffusion in the Node of Ranvier by Mesh Adaptation. *PLoS ONE* **11**, e0161318 (2016).

33. Lopreore, C. L. *et al.* Computational modeling of three-dimensional electrodiffusion in biological systems: application to the node of Ranvier. *Biophysical journal* **95**, 2624–2635 (2008).

34. Gulati, R. & Rudraraju, S. Spatio-temporal modeling of saltatory conduction in neurons using Poisson–Nernst–Planck treatment and estimation of conduction velocity. *Brain Multiphysics* **4**, 100061 (2023).

35. Dione, I., Doyon, N. & Deteix, J. Sensitivity analysis of the Poisson Nernst–Planck equations: a finite element approximation for the sensitive analysis of an electrodiffusion model. *J. Math. Biol.* **78**, 21–56 (2019).

36. Hanemaaijer, N. A. *et al.* Ca2+ entry through NaV channels generates submillisecond axonal Ca2+ signaling. *eLife* **9**, e54566 (2020).

37. Battefeld, A., Tran, B. T., Gavrilis, J., Cooper, E. C. & Kole, M. H. P. Heteromeric Kv7.2/7.3 channels differentially regulate action potential initiation and conduction in neocortical myelinated axons. *J Neurosci* **34**, 3719–3732 (2014).

38. Pryor, R. W. *Multiphysics Modeling Using COMSOL®: A First Principles Approach*. (Jones & Bartlett Learning, 2011).

39. Savtchenko, L. P. *et al.* Disentangling astroglial physiology with a realistic cell model in silico. *Nat. Commun.* **9**, 3554 (2018).

40. Sibille, J., Duc, K. D., Holcman, D. & Rouach, N. The Neuroglial Potassium Cycle during Neurotransmission: Role of Kir4.1 Channels. *PLoS Comput. Biol.* **11**, e1004137 (2015).

41. Hallermann, S., Kock, C. P. J. de, Stuart, G. J. & Kole, M. H. P. State and location dependence of action potential metabolic cost in cortical pyramidal neurons. *Nature Neuroscience* **15**, 1007–1014 (2012).
42. Schmidt-Hieber, C. & Bischofberger, J. Fast sodium channel gating supports localized and efficient axonal action potential initiation. *J. Neurosci.* **30**, 10233–10242 (2010).
43. Peters, A. & Sethares, C. Is there remyelination during aging of the primate central nervous system? *J. Comp. Neurol.* **460**, 238–254 (2003).
44. Battefeld, A., Klooster, J. & Kole, M. H. P. Myelinating satellite oligodendrocytes are integrated in a glial syncytium constraining neuronal high-frequency activity. *Nature communications* **7**, 11298 (2016).
45. Brohawn, S. G. *et al.* The mechanosensitive ion channel TRAAK is localized to the mammalian node of Ranvier. *eLife* **8**, (2019).
46. Frankenhaeuser, B. & Huxley, A. F. The action potential in the myelinated nerve fibre of Xenopus laevis as computed on the basis of voltage clamp data. *J Physiology* **171**, 302–315 (1964).
47. Cullen, C. L. *et al.* Periaxonal and nodal plasticities modulate action potential conduction in the adult mouse brain. *Cell Reports* **34**, 108641 (2021).
48. Savtchenko, L. P. & Rusakov, D. A. The optimal height of the synaptic cleft. *Proceedings of the National Academy of Sciences of the United States of America* **104**, 1823–1828 (2007).
49. Tyurikova, O. *et al.* Astrocyte Kir4.1 expression level territorially controls excitatory transmission in the brain. *Cell Rep.* **44**, 115299 (2025).
50. Nans, A., Einheber, S., Salzer, J. L. & Stokes, D. L. Electron tomography of paranodal septate-like junctions and the associated axonal and glial cytoskeletons in the central nervous system. *Journal of Neuroscience Research* **89**, 310–319 (2010).
51. Schloetter, M., Maret, G. U. & Kleineidam, C. J. Annihilation of action potentials induces electrical coupling between neurons. *eLife* **12**, RP88335 (2025).
52. Katz, B. & Schmitt, O. H. Electric interaction between two adjacent nerve fibres. *J. Physiol.* **97**, 471–488 (1940).
53. Labarchède, M., Petrel, M. & Battefeld, A. Neuronal activity induces myelin voltage changes that reflect action potential dependent myelin potassium buffering. *bioRxiv* 2025.12.16.694668 (2025) doi:10.64898/2025.12.16.694668.
54. Larson, V. A. *et al.* Oligodendrocytes control potassium accumulation in white matter and seizure susceptibility. *eLife* **7**, (2018).
55. Rosenbluth, J. Multiple functions of the paranodal junction of myelinated nerve fibers. *J. Neurosci. Res.* **87**, 3250–3258 (2009).
56. Poliak, S. & Peles, E. The local differentiation of myelinated axons at nodes of Ranvier. *Nature Reviews Neuroscience* **4**, 968–980 (2003).
57. Kozar-Gillan, N. *et al.* LGI3/2–ADAM23 interactions cluster Kv1 channels in myelinated axons to regulate refractory period. *J. Cell Biol.* **222**, e202211031 (2023).
58. Eccles, J. C. & Jaeger, J. C. The relationship between the mode of operation and the dimensions of the junctional regions at synapses and motor end-organs. *Proc. R. Soc. Lond. Ser. B - Biol. Sci.* **148**, 38–56 (1958).
59. Nave, K.-A. Myelination and the trophic support of long axons. *Nature Reviews Neuroscience* **11**, 275–283 (2010).
60. Rey, S., Zalc, B. & Klämbt, C. Evolution of glial wrapping: A new hypothesis. *Dev. Neurobiol.* **81**, 453–463 (2021).
61. Pan, Z. *et al.* A common ankyrin-G-based mechanism retains KCNQ and NaV channels at electrically active domains of the axon. *J. Neurosci.* **26**, 2599–2613 (2006).

62. Kanda, H. *et al.* TREK-1 and TRAAK Are Principal K+ Channels at the Nodes of Ranvier for Rapid Action Potential Conduction on Mammalian Myelinated Afferent Nerves. *Neuron* **104**, 960-971.e7 (2019).

63. Jr., G. E., Wu, Y., Ogawa, Y., Ding, X. & Rasband, M. N. An evolutionarily conserved AnkyrinG-dependent motif clusters axonal K2P K+ channels. *J. Cell Biol.* **223**, e202401140 (2024).

64. Brohawn, S. G., Su, Z. & MacKinnon, R. Mechanosensitivity is mediated directly by the lipid membrane in TRAAK and TREK1 K+ channels. *Proc. Natl. Acad. Sci.* **111**, 3614–3619 (2014).

65. Saeedimasine, M., Montanino, A., Kleiven, S. & Villa, A. Elucidating Axonal Injuries Through Molecular Modelling of Myelin Sheaths and Nodes of Ranvier. *Front. Mol. Biosci.* **8**, 669897 (2021).

66. Gentet, L. J., Stuart, G. J. & Clements, J. D. Direct measurement of specific membrane capacitance in neurons. *Biophysical journal* **79**, 314–320 (2000).

67. Singer, S. J. & Nicolson, G. L. The fluid mosaic model of the structure of cell membranes. *Science* **175**, 720–731 (1972).

68. Chan, C.-F. *et al.* Ba2+- and bupivacaine-sensitive background K+ conductances mediate rapid EPSP attenuation in oligodendrocyte precursor cells. **591**, 4843–4858 (2013).

69. Kushmerick, M. J. & Podolsky, R. J. Ionic Mobility in Muscle Cells. *Science* **166**, 1297–1298 (1969).

**Acknowledgements**

The authors thank Dr. Predrag Janjic for critical reading of the manuscript.

**Data availability**

Data will be made openly available upon publication.

## Online Methods

*FE simulation details*

With the FE software package COMSOL [37] we study an azimuthally symmetric axonal geometry with longitudinal coordinate $z$ and radial coordinate $r$, schematically illustrated in **Fig. 1a-c**. Inside the modelled compartments we consider ionic concentration $\rho_i(r, z, t)$ for species $i$ and the electric potential $\Psi(r, z, t)$. For compactness we will omit explicitly writing $r$, $z$ or $t$ in the argument when they are not needed. The compartments contain aqueous electrolytes with electric permittivity $\epsilon = 0.71\ \mathrm{nF \cdot m^{-1}}$. The electrolyte contains $Na^+$, $K^+$, and a generic monovalent anion for charge neutrality. At the far end of the extracellular space, we impose a ground and extracellular bulk ion concentrations $\rho_i = \rho_{i,\mathrm{b}}$. To obtain a steady-state solution from which we initialize the time-dependent simulation, intracellular ionic concentrations are also imposed, while for time-dependent studies this condition is removed and (nearly) all intracellular concentrations can accumulate or deplete. An exception to this is that the concentrations and voltages remain fixed inside the OL cytoplasm. Our model does not include the further complex route of charged ions to move via gap junctions, and astrocytic Kir channels, connecting the OL cytoplasm ultimately to the blood circulation [18–21] . These are now coarsely approximated by fixed internal voltages and concentrations, as indicated in **Supplementary Fig. 1**. Without a fixed value the AP-induced $K^+$ influx rises the internal OL cytoplasm voltage to a plateau without returning to resting potential as there is no charge efflux route.

For the interiors of the axonal geometry transport is described by the Poisson-Nernst-Planck (PNP) Eqs. (1)-(3) given by

$$\nabla^2 \Psi = -\frac{e}{\epsilon} \sum_i z_i \rho_i, \tag{1}$$

$$\frac{\partial \rho_i}{\partial t} + \nabla \cdot \mathbf{j}_i = 0, \tag{2}$$

$$\mathbf{j}_i = -D \left( \nabla \rho_i + z_i \rho_i \frac{e \nabla \Psi}{k_{\mathrm{B}} T} \right). \tag{3}$$

Electrostatics is accounted for by the Poisson Eq. (1), the conservation of ions by the continuity Eq. (2) with ionic fluxes $\mathbf{j}_i$, and Fickian diffusion and Ohmic conduction by the Nernst-Planck Eq. (3).

Ionic currents through the membrane are implemented by established Hodgkin-Huxley-like channel equations (more detail below). Fully resolving the physics inside the membrane requires simulations on the much smaller molecular or atomic scale, for which the PNP equations are not suited. This does mean that the modelled ionic fluxes through the membrane are not driven by the local electric field, which in principle is known within the model, but rather by the overall voltage drop $V_j$ over the membrane j. We are considering multiple neighboring membranes between the axon core and the grounded extracellular space, thus we define $V_j$ as the difference in voltage between the middle points of two compartments at either side of the membrane, i.e. $V_j(z,t) = \Psi(r_{j-}, z, t) - \Psi(r_{j+}, z, t)$. Here, $j-$ $(j+)$ refers to the inner (outer) compartment, with $r_{j\pm}$ the radial coordinate in the middle of the respective compartment.

*Ion channel models*

To implement the membrane currents as described above, we apply flux boundary conditions on the boundaries of the form

$$e\mathbf{n}_j \cdot \mathbf{j}_i = g_i(t)\left(V_j(t) - E_{i,j}(t)\right) + C_j\dot{V}_j(t), \tag{4}$$

with $\mathbf{n}_j$ membrane $j$'s inward normal vector, $g_i(t)$the membrane conductance (per unit area) of ionic species $i$, $V_j$the voltage over membrane $j$, $E_{i,j}(z,t) = \frac{k_B T}{e}\ln\left(\rho_i(r_{j+},z,t)/\rho_i(r_{j-},z,t)\right)$ the reversal potential of ion $i$ over membrane $j$, $C_j$ the capacitance of membrane $j$, and $e$ is the elementary charge. For the axolemma and adaxonal membrane (neuron and oligodendrocyte, respectively) a biophysically realistic value $C_m = C_{ad} = 1\ \mu\text{F cm}^{-2}$ was chosen [7,62] and for the compact myelin $C_{my} = 0.04\ \mu\text{F cm}^{-2}$, corresponding to ~13 myelin lamellae [7].

The (continuum) electrodiffusion equations we use here do not feature the atomic level resolution required to solve for ion transport inside the membrane channels. Instead, we incorporated established Hodgkin-Huxley-type channel equations with dynamically updated reversal potentials to impose quantitatively accurate ionic fluxes across the membrane surfaces, without having to solve for the inside of the actual membrane itself. In order to describe $g_i(t)$, we implement two types of voltage-gated ion channels; the 8-state fast K$v$1-type channel model [40], with channel parameters as in Ref. [7] , and the $Na^+$ channel based on axonal recording Ref. [42]. K$v$1 channels are placed at the juxtaparanode with a maximum conductance density of $g_K = 6$ nS·$\mu$m$^{-2}$ and the voltage-gated $Na^+$ channels were exclusively at the node of Ranvier with a peak conductance density $g_{Na} = 30$ nS·$\mu$m$^{-2}$ (**Fig. 1c**, **Supplementary Table 1**). Nodal Kv7 channels were omitted as with the slow time course of activation they do not directly contribute to nodal AP waveform but instead set the resting potential and determine $Na^+$ channel availability [37]. All conductance densities are within range of earlier cable-based models [7,26,41].

A mathematical model of Kir4.1 channels [38,39] is placed in the adaxonal membrane separating the juxtaparanodal periaxonal space from the inner cytoplasmic tongue. As Kir4.1 plays a central role in the results in this work, we repeat the governing equation that models the Kir4.1 conductance here as well. Specifically, the channel description was based on the Kir4.1 model used for astrocyte nanoscale simulations [49]. The current density $I_{Kir}$ through Kir4.1 channels is approximated by

$$I_{\mathrm{Kir}} = \overline{g}_{\mathrm{Kir}}(V_{\mathrm{ad}} - V_{\mathrm{KA}} - V_{\mathrm{A1}})\left(\frac{\sqrt{\rho_{\mathrm{K,o}}/\rho_{\mathrm{ref}}}}{1+\exp\left(\frac{V_{\mathrm{ad}}-V_{\mathrm{KA}}-V_{\mathrm{A2}}}{V_{\mathrm{A3}}}\right)}\right). \tag{5}$$

Here $V_{\mathrm{ad}}$ is the adaxonal membrane potential, $V_{\mathrm{KA}} = N_{\mathrm{K}} E_{\mathrm{K^+,\,ad}}$ where $N_{\mathrm{K}} = 0.81$ a normalisation factor, $V_{\mathrm{A1}} = -16.33$ mV such that $I_{\mathrm{Kir}} = 0$ if $V_{\mathrm{ad}} = E_{\mathrm{K^+,\,ad}}$, $\rho_{\mathrm{K,o}}$ is the $\mathrm{K^+}$concentration at the middle of the periaxonal space, $\rho_{\mathrm{ref}} = 1$ mM is a reference concentration to fix the units, $V_{\mathrm{A2}} = 34$ mV, and $V_{\mathrm{A3}} = 19.23$ mV. Kir4.1 channels are placed in the adaxonal membrane with a channel density of $\overline{g}_{\mathrm{Kir}} = 5$ nS $\mu\mathrm{m}^{-2}$. Note that $E_{\mathrm{K^+,\,ad}}$ is updated dynamically according to the dynamic $\mathrm{K^+}$ concentrations, as laid out below Eq. (4). In addition, $\mathrm{Na^+}$ and $\mathrm{K^+}$ leak channels and capacitive currents were incorporated with typical physiological capacitances for both the axolemma and the compact myelin (see **Supplementary Table 1**).

*Swelling estimate from $K^+$ accumulation*

To assess whether $\mathrm{K^+}$ accumulation could induce mechanical deformation of the periaxonal space, we estimate the resultant pressure and corresponding height increase. The osmotic pressure induced by a concentration difference $\Delta\rho_{\mathrm{K^+}}$ follows the van 't Hoff relation

$$\Pi = \Delta\rho_{\mathrm{K^+}} k_{\mathrm{B}} T.$$

For a persistent $\mathrm{K^+}$ excess of $\sim 10$ mM, this yields $\Pi \approx 26$ kPa at body temperature (310 K). For a cylindrical membrane of radius $R$, the Young-Laplace equation relates this osmotic pressure difference $\Pi$ to circumferential wall tension $\gamma$

$$\gamma = \Pi \cdot R.$$

Coupling this with Hooke's law for elastic stretching, $\gamma = K_{\mathrm{A}}\varepsilon = K_{\mathrm{A}}\Delta h/R$ with stiffness parameter $K_{\mathrm{A}}$, we obtain the equilibrium height increase

$$\Delta h = \frac{\Pi\, R^2}{K_{\mathrm{A}}}.$$

Estimates for $K_{\mathrm{A}}$ vary but are mostly comparable [63,64] (Saeedimasine et al., 2019). Using the axon radius $R = 0.5\,\mu$m and estimating $K_{\mathrm{A}} = 0.34$ N/m [65] for a single myelin layer or $K_{\mathrm{A}} \sim 4.4$ N/m for the full myelin stack (~13 lamellae acting in parallel), the predicted swelling for a sustained 10 mM $\mathrm{K^+}$ elevation would be $\sim 1.5$ nm respectively. Thus, while the osmotic pressure is substantial relative to tissue moduli, the multilamellar architecture of myelin provides sufficient mechanical rigidity to limit swelling to sub-nanometre scales when $K^+$ is appropriately cleared. However, without $K^+$ clearing our model suggests $K^+$ increases of several tens of mM, due to which the nanoscopic periaxonal space could substantially swell. Additionally, for a single membrane the swelling would be $\sim 19$ nm for a sustained 10 mM $K^+$ elevation, if we take the same $K_{\mathrm{A}} \sim 0.34$ N/m modulus for simplicity.

**Supplementary Data**

**Supplementary Table 1. Parameters of the FE model**

| Parameter | Symbol | Value | Reference |
|---|---|---|---|
| $Na^+$ conductance density (maximum) | $\overline{g}_{Na}$ | 30 nS·$\mu m^{-2}$ | [7,26] |
| K$v$1 conductance density | $\overline{g}_{K}$ | 6.0 nS·$\mu m^{-2}$ | [41] |
| Kir4.1 conductance density | $\overline{g}_{Kir}$ | 5.0 nS·$\mu m^{-2}$ | |
| $Na^+$ leak conductance density | $\overline{g}_{l,Na}$ | 0.4 pS·$\mu m^{-2}$ | |
| $K^+$ leak conductance density | $\overline{g}_{l,K}$ | 4.0 pS·$\mu m^{-2}$ | |
| $K^+$ leak conductance density adaxonal membrane | $\overline{g}_{ad}$ | 2.4 pS·$\mu m^{-2}$ | |
| Myelin sheath resistance (13 lamellae) | $R_{my}$ | 240 kΩ·$cm^2$ | [7] |
| Axolemma and adaxonal membrane capacitance | $C_m$, $C_{ad}$ | 1.0 $\mu$F·$cm^{-2}$ | [7,66] |
| Compact myelin capacitance | $C_{my}$ | 0.04 $\mu$F·$cm^{-2}$ | [7] |
| | | | |
| Internode length [a] | | 2.0 $\mu$m | |
| Juxtaparanode (JXP) length | | 3.0 $\mu$m | |
| Paranode length | | 1.5 $\mu$m | |
| Node length [b] | | 2.0 $\mu$m | [26,28] |
| Axon radius | $r$ | 0.5 $\mu$m | |
| Axolemma height | | 8.0 nm | [67] |
| Periaxonal space height | $\delta_{pa}$ | 12.0 nm | [7] |
| Paranode height | $\delta_{pn}$ | 7.4 nm | [50] |
| Cytoplasmic tongue height | | 50 nm | [50] |
| | | | |
| Extracellular $Na^+$ concentration [c] | $[Na^+]_o$ | 145 mM | |
| Intracellular $Na^+$ concentration (axon and tongue) | $[Na^+]_i$ | 12 mM | |
| Extracellular $K^+$ concentration | $[K^+]_o$ | 4 mM | |
| Intracellular $K^+$ concentration (axon) | $[K^+]_i$ | 155 mM | |
| Intracellular $K^+$ concentration (tongue) [d] | | 120 mM | [68] |
| | | | |
| Diffusion coefficient $Na^+$ | $D_{Na}$ | 0.6 $\mu m^2$·$s^{-1}$ | [69] |
| Diffusion coefficient $K^+$ | $D_K$ | 1.0 $\mu m^2$·$s^{-1}$ | [69] |
| Diffusion coefficient generic anion | $D$- | 1.0 $\mu m^2$·$s^{-1}$ | |
| | | | |

[a] Only small fraction of the internode is included for computational efficiency.

[b] Upper end value of layer 5 pyramidal neuron node lengths.

[c] All concentrations are initialization values.

[d] Concentration set to obtain a $V_{ad}$ of approximately −86 mV at rest [44].

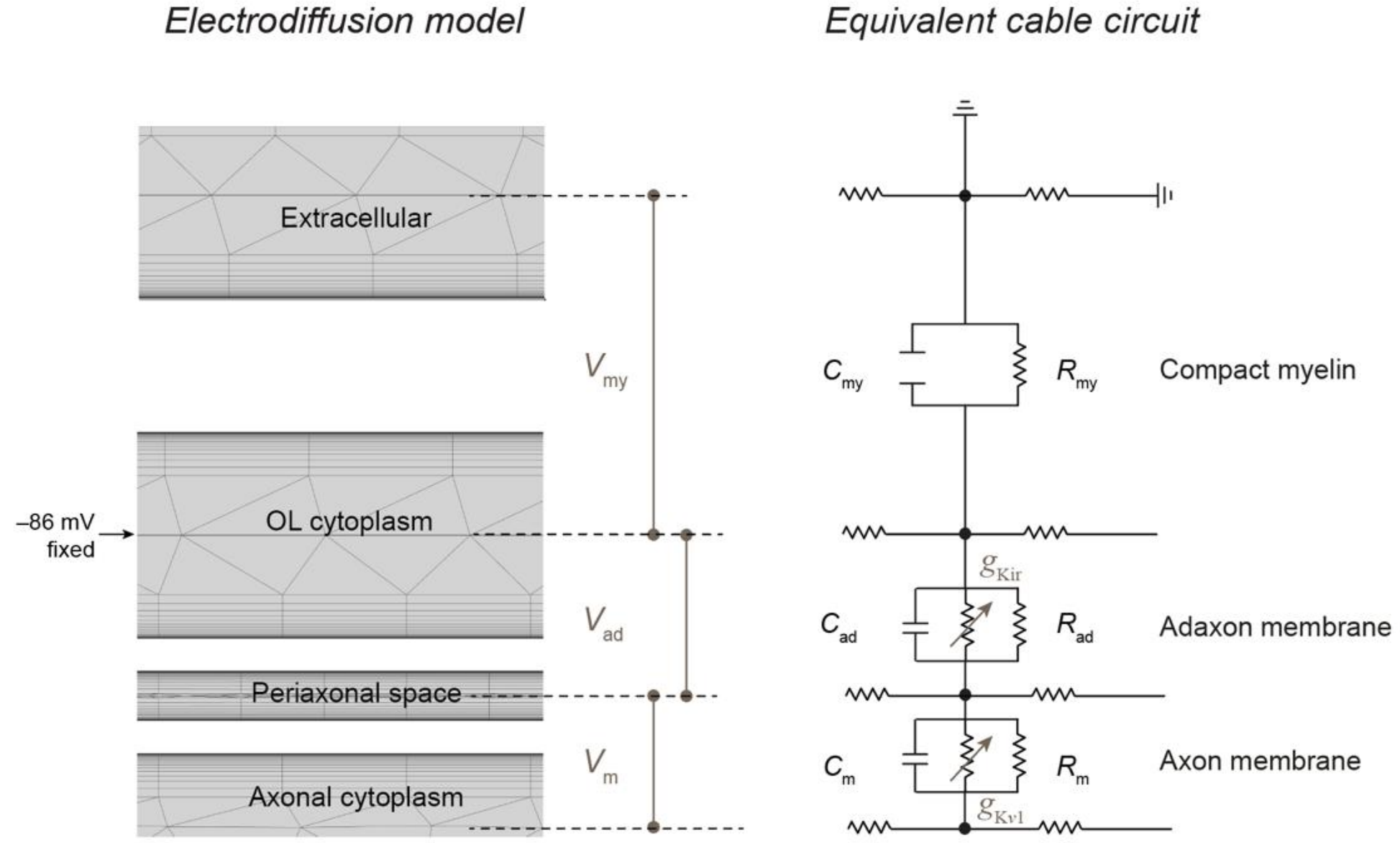

**Supplementary Fig. 1 | Voltage division in the FE model**

Left, zoom of the 2D mesh (gray) at the juxtaparanode compartment used in the finite element model. Right, approximate equivalent cable model for reference. Note that all results in this work result from the solving the PNP equations within the finite element model, the circuit is depicted for explanatory reasons. Similar to a triple cable resistor-capacitor ($RC$) model, there are three distinct longitudinal current pathways in addition to extracellular ground. The voltage and ionic concentrations at the middle of the OL cytoplasm are fixed. This results in two separate membrane voltages $V_m$ and $V_{ad}$, in which voltage-gated $K^+$ channels are embedded, axonal K$v$1 and myelinic Kir4.1 in this model, respectively.

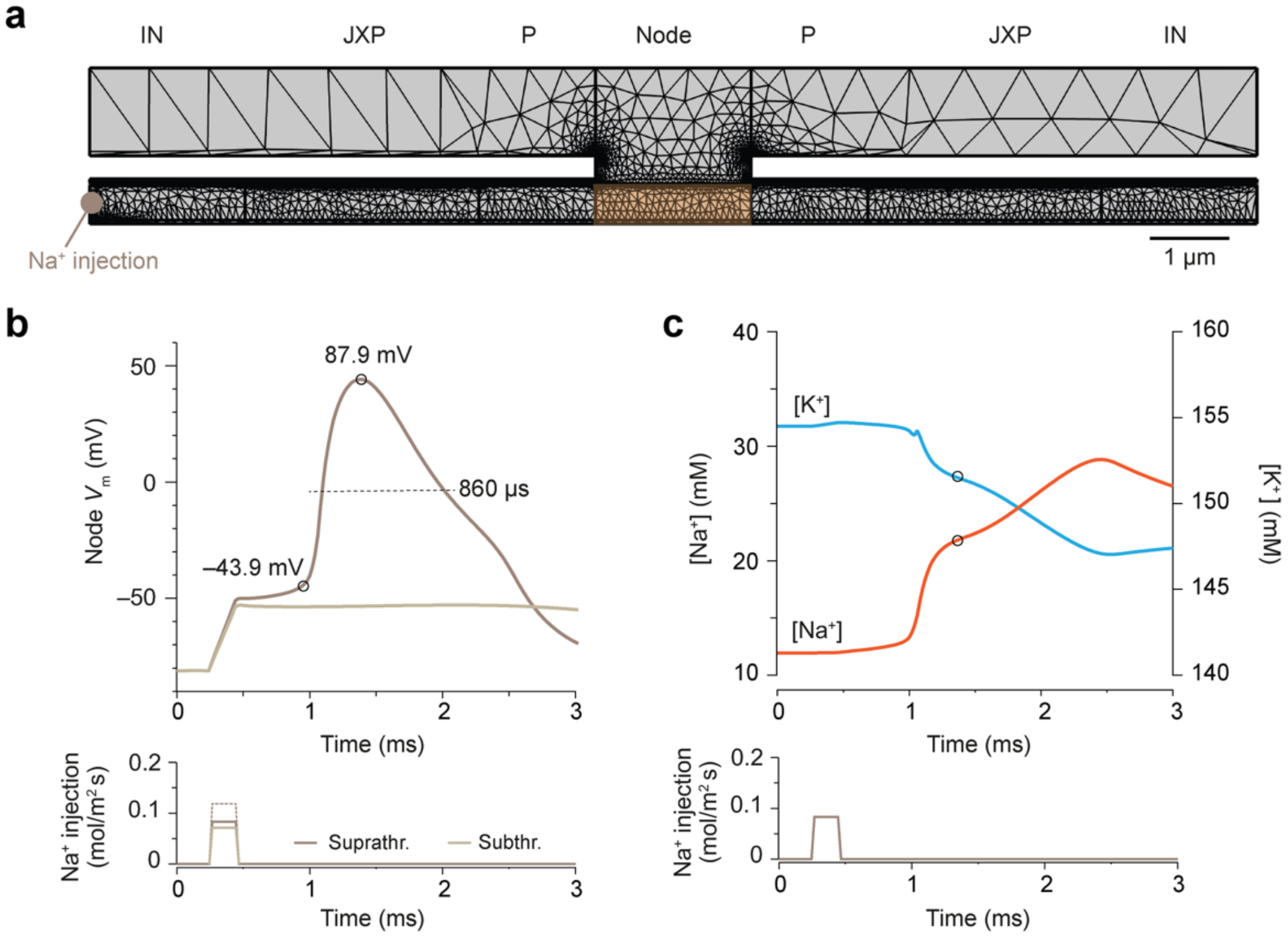

**Supplementary Fig. 2 | Ionic current influx during the nodal AP**

(**a**) 2D overview of the complete 2D mesh structure in the axon and extracellular space across continuous sections representing internodes (IN), juxtaparanodes (JXP), paranodes (P) and the node (N). Total 3D volume of the intracellular nodal space of the PNP simulations is 1.57 $\mu m^3$. To evoke APs a brief $Na^+$ current step is injected inside the axoplasm at the edge of the model. (**b**) Top, example voltage-time plots of nodal $V_m$ evoked by a subthreshold (light brown) $Na^+$ injection and a slightly larger $Na^+$ injection, triggering a full AP (dark brown). Note the transition from subthreshold $V_m$ into opening of $Na^+$ channels at a threshold of –43.9 mV, causing axonal $Na^+$ influx. The AP rapidly overshoots to a peak depolarization of +46 mV, reflecting an AP with 87.9 mV amplitude and 860 µs halfwidth. Dotted $Na^+$ injection is a higher injection used for the standard AP simulations in the simulations. (**c**) Overlay of the axoplasmic $Na^+$ (red) and $K^+$ (blue) concentrations in the nodal versus time. In the absence of nodal repolarizing conductances the voltage-mediated closing of $Na^+$ channels during repolarization is incomplete and causing excess $Na^+$ influx due to the increasing driving force. See also **Supplementary Movie 1**.

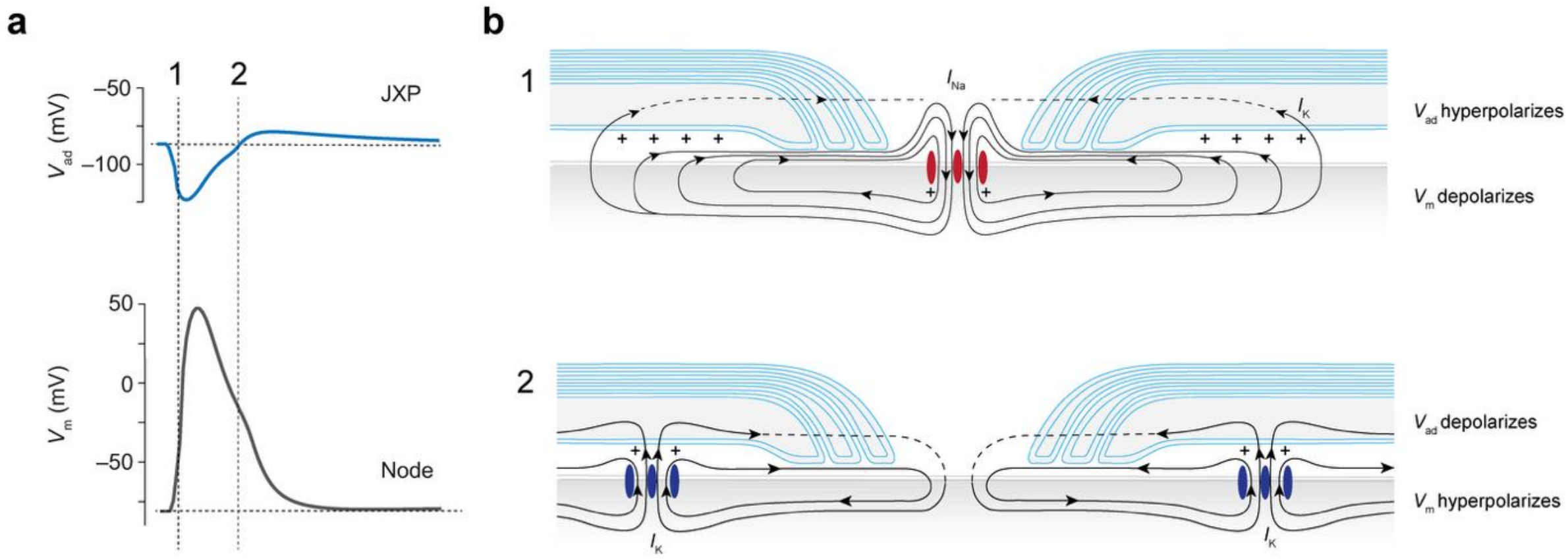

**Supplementary Fig. 3│ Local circuit currents at the node of Ranvier**
(**a**) Nodal AP with the $Na^+$ current mediated rising phase (1) and falling phase (2), temporally aligned with the hyperpolarizing adaxonal action potential. (**b**) Schematic representation of the circuit loop mediated in the first hundred microseconds of the nodal AP (1) associated with the $Na^+$ influx. Positive charging of the axoplasm is associated with bidirectional axial charge redistribution and returning current loops back to the extracellular space. Returning current loops flow via the paranodal and periaxonal spaces as well as the inner myelin tongue. Capacitive currents and $K^+$ will leave the myelin sheath via gap junctions and astrocytes (dotted line, not simulated in the FE model). During the falling phase (2) the current loops from JXP $K^+$ efflux in the periaxonal space return to ground via the extracellular potentials.